\documentclass[aps,preprint,nofootinbib,showpacs,superscriptaddress,showkeys]{revtex4-1}
\date{\today}
\usepackage{graphicx}
\usepackage{color}
\usepackage{amsmath}
\usepackage{amsfonts}
\usepackage{amssymb}

\begin{document}

\title{Analytic proof of partial conservation of seniority in $j=9/2$ shells}
\author{Chong Qi}
\email{chongq@kth.se}
\affiliation{KTH (Royal Institute of Technology), Alba Nova University Center,
SE-10691 Stockholm, Sweden}
\author{Z.X. Xu}
\affiliation{KTH (Royal Institute of Technology), Alba Nova University Center,
SE-10691 Stockholm, Sweden}
\author{R.J. Liotta}
\affiliation{KTH (Royal Institute of Technology), Alba Nova University Center,
SE-10691 Stockholm, Sweden}

\begin{abstract}
A partial conservation of the seniority quantum number in
$j=9/2$ shells has been found recently in a numerical application.
In this paper a complete analytic proof for this problem is derived as an extension of the work by Zamick and P. Van Isacker [Phys. Rev. C 78 (2008) 044327].
We analyze the properties of the non-diagonal matrix elements with the help of the one-particle and two-particle coefficients of
fractional parentage (cfp's).
It is found that all non-diagonal (and the relevant diagonal) matrix elements can be re-expressed in simple ways and are proportional to certain one-particle cfp's.
This
remarkable occurrence of partial dynamic symmetry is
the consequence of the peculiar property of the $j=9/2$ shell, where all $v=3$ and 5 states are uniquely defined.
\end{abstract}

%\pacs{21.60.Cs, 03.65.Fd, 27.60.+j, 21.30.Fe}

\maketitle
\section{Introduction}
The concept of seniority quantum number in many-body systems has played a very important role since its inception by Racah \cite{Racah43}.
It refers to the minimum number of unpaired particles in a single-$j$ shell for a given configuration $|j^n;I\rangle$ where $I$ is the total angular momentum. In nuclear physics it has classified the influence of the pairing force on nuclear spectra \cite{Shalit63,Talmi93}. But this concept is nowadays been applied in a variety of fields, including Bose-Einstein condensates \cite{isac10}.
It is established that seniority is a good quantum number for
systems with identical fermions in shells with
$j\leq7/2$. All states in these systems can be uniquely specified by the total angular momentum $I$ and seniority $v$.
Unfortunately seniority symmetry breaks in
shells with $j\geq9/2$. Efforts have been made to find cases for which that symmetry is partially fulfilled. It is thus found that the rotationally-invariant interaction has to satisfy a
number of constraints in order to conserve seniority~\cite{Talmi93}.  The conservation conditions are not satisfied by most general
two-body interactions for which the eigenstates would be admixtures
of states with different seniorities. However, it was noted that in
$j=9/2$ shell two special eigenstates with $I=4$ and 6 have good seniority for an
arbitrary interaction~\cite{Escuderos06,Zamick07}. The states are
eigenstates of any spherically symmetric two-body interaction. They exhibit partial dynamic
symmetry and the solvability property (for details, see, e.g., Refs.~\cite{Isacker08,Levi10}). 

The problem has been described in Refs. \cite{Escuderos06,Zamick07,Isacker08,Zam08,Levi10,Qi11} in a variety of ways and will only be briefly presented here for completeness. For a system with $n=4$ identical fermions in a $j=9/2$ shell, The $I=4$ (and $I=6$) states may be constructed so that one state has seniority $v=2$  (denoted as
$|j^4,v=2,I\rangle$ in the following) and the other two have seniority $v=4$ (denoted as
$|j^4,\alpha_1,v=4,I\rangle$ and $|j^4,\alpha_2,v=4,I\rangle$ where the index $\alpha$ symbolizes
any additional quantum number needed when there are more than one state with a given seniority $v$ and total angular momentum $I$.
The seniority $v=4$ states are not uniquely defined and any linear combination
of them would result in a new sets of $v=4$ states.
The corresponding Hamiltonian matrix elements can be written as linear combinations
of the interaction terms $V_J=\langle j^2;J|\hat{V}|j^2;J\rangle$ as,
\begin{equation}
H^I_{ij}=n(n-1)/2\sum_JM^I_{ij}(J)V_J,
\end{equation}
where the angular momenta $J$ can take even values between 0 and $2j-1$. The symmetric matrices $M^I(J)$ can be constructed with the help of one-particle or two-particle coefficients of factional parentage (cfp's).  For a seniority-conserving interaction it is $H^I_{2\alpha_1}=H^I_{2\alpha_2}=0$. The matrix elements of $M^I(J)$ do not vanish in general. However, in Refs.~\cite{Escuderos06,Zamick07} it was found that one special $v=4$, $I=4$ and $I=6$ (denoted as $|a\rangle$ in Ref.  \cite{Zam08})
 state has the interesting property that it has vanishing matrix elements with the remaining $v=2$ and 4
 states orthogonal to $|a\rangle$ even if an interaction which does
not conserve seniority is used. This indicates that one should have $M^I_{2a}(J)=M^I_{ab}(J)=0$ for all $J$ values, where $|b\rangle$ denotes the $v=4$ state orthogonal to the state $|a\rangle$. 

The
consequences of the vanishing of non-diagonal matrix elements was examined in Ref. \cite{Zam08} for the cases of $I=J$. It was thus shown 
that this is due to a special relation of certain
one-particle cfp's, namely
\begin{eqnarray}\label{cfp1}
\nonumber \frac{[j^4(\alpha_1,v=4,I)jI_5|\}j^5,v=3,I_5=j]}{[j^4(\alpha_2,v=4,I)jI_5|\}j^5,v=3,I_5=j]} \\
\nonumber=\frac{[j^4(\alpha_1,v=4,I)jI_5|\}j^5,v=5,I_5=j]}{[j^4(\alpha_2,v=4,I)jI_5|\}j^5,v=5,I_5=j]}\\
=\frac{[j^3 (\nu_3=3, I_3=j)j I |\}j^{4},\alpha_1,\nu=4, I] }{[j^3 (\nu_3=3, I_3=j)j I |\}j^{4},\alpha_2,\nu=4, I] },
\end{eqnarray}
where the $v=3$ and 5 states 
are uniquely specified by their angular momenta $I$ and seniority $v$. By taking this fact into account, an analytic proof can be derived for above special relation
through the principal-parent procedure~\cite{Qi11}. 
The objective of this work is to derive a simple and complete proof of
the vanishing of non-diagonal matrix elements of $M^I(J)$ for all $J$ values. To do so we will explore further the special relations of one-particle cfp's based on the variety of recursion relations proposed in Refs. \cite{Shalit63,Sun89,Zam06}. We will show that all matrix element of the matrices $M^I(J)$ can be re-expressed in simple ways. This subject will also be analyzed in terms of two-particle cfp's. 

\section{Recursion relations}

The $v\rightarrow v-1$ one-particle cfp's can be factorized as \cite{Sun89}\footnote{For clearer presentation, the notations used here are slightly different from those of Refs. \cite{Sun89,Qi11}.}
\begin{eqnarray}\label{sun}
%\nonumber 
[j^{n-1}(\alpha_1,v-1,J_1)jJ|\}j^{n}\alpha v J]
=\sqrt{\frac{v(2j+3-n-v)}{n(2j+3-2v)}}
 R(j,v-1,\alpha_1J_1;v\alpha J).
\end{eqnarray}
The factor $R$ in this equation can be constructed with the principal-parent procedure as
\begin{eqnarray}
 R(j,v-1,\alpha_1J_1;v\alpha J) =
 \sum_{\alpha_1'J_1'}c_{\alpha_1'J_1'}
 R'(j,v-1,\alpha_1J_1;v(\alpha_1'J_1') J),
\end{eqnarray}
where $\alpha_1'J_1'$ denote the principal parents. The coefficients $c_{\alpha_1'J_1'}$ can be determined by the standard orthonormalization procedure. For a state that can be uniquely defined by the quantum numbers $v$ and $J$, it can be constructed equivalently by taking different principal parents and one has $c_{\alpha_1'J_1'}=1$.
The isoscalar factor $R'$ can be calculated using the recursion relation given by,
\begin{equation}
R(j,v-1,\alpha_1J_1;v(\alpha_1'J_1') J)
=\frac{P(\alpha_1'J_1'\alpha_1J_1J)}{\sqrt{v P(\alpha_1'
J_1'\alpha_1'J_1'J)}},
\end{equation}
where
\begin{eqnarray}
\nonumber P(\alpha_1'J_1'\alpha_1J_1J) =
\delta_{\alpha_1'\alpha_1}\delta_{J_1'J_1} +
\nonumber (-1)^{J+J_1'}(v-1)\sqrt{(2J_1'+1)(2J_1+1)} \sum_{\alpha_2
J_2}\left[\left\{
            \begin{array}{ccc}
              j & J_2 & J_1' \\
              j & J & J_1 \\
            \end{array}
          \right\}\right.\\
        \left.  +\frac{(-1)^{v}2\delta_{J_2J}}{(2J+1)(2j+5-2v)}\right] R(j,v-2,\alpha_2J_2;v-1,\alpha_1 J_1)R(j,v-2,\alpha_2J_2;v-1\alpha_1' J_1').
\end{eqnarray}

From these equations one finds that the following relation should hold,
\begin{eqnarray}
 R(j,v-1,\alpha_1J_1;v(\alpha_1'J_1') J)
R(j,v-1,\alpha_1'J_1';v(\alpha_1'J_1') J)
=\frac{P(\alpha_1'J_1'\alpha_1J_1J)}{v}.
\end{eqnarray}
For the special case of $n=v$ of concern in this work, we simply have
\begin{eqnarray}
 [j^{n-1}(\alpha_1,v-1,J_1)jJ|\}j^{n}\alpha v J]=
 R(j,v-1,\alpha_1J_1;v\alpha J),~~~
\end{eqnarray}
and
%\begin{widetext}
\begin{eqnarray}
\nonumber &&v[j^{n-1}(\alpha_1,v-1,J_1)jJ|\}j^{n} v (\alpha_1'J_1') J][j^{n-1}(\alpha_1',v-1,J_1')jJ|\}j^{n} v (\alpha_1'J_1') J]\\
\nonumber &=&\delta_{\alpha_1'\alpha_1}\delta_{J_1'J_1} + (-1)^{J_1+J_1'}(v-1)\sqrt{(2J_1'+1)(2J_1+1)}
\sum_{\alpha_2
J_2}\left[\left\{
            \begin{array}{ccc}
             J_2 &  j & J_1' \\
            J &  j &  J_1 \\
            \end{array}
          \right\}
          +\frac{(-1)^{v}2\delta_{J_2J}}{(2J+1)(2j+5-2v)}\right]\nonumber\\
&& \times [j^{n-2}(\alpha_2,v-2,J_2)jJ_1|\}j^{n-1},\alpha_1, v-1, J_1] [j^{n-2}(\alpha_2,v-2,J_2)jJ_1'|\}j^{n-1},\alpha_1', v-1, J_1'].~~
\end{eqnarray}
%\end{widetext}
This relation is similar to the Redmond recursion relation \cite{Red54,Talmi93} and the modified Redmond relation used in Refs. \cite{Zam06,Zam08}. The difference between this recursion relation and those of Refs. \cite{Red54,Zam06} is that the term in the left-hand side of Eq. (9) have fixed seniority quantum number. Moreover, the state $|j^nv(\alpha_1'J_1')J\rangle$ constructed through the principal parent procedure can always be written as an expansion of the basis set $|\beta vJ\rangle$ with quantum numbers $v$ and $J$ as,
\begin{eqnarray}
[j^{n-1}(\alpha_1,v-1,J_1)jJ|\}j^{n} v (\alpha_1'J_1') J]=\sum_{\beta_i}\beta_i[j^{n-1}(\alpha_1,v-1,J_1)jJ|\}j^{n}\beta_i v J],
\end{eqnarray}
where $\beta_i$ denotes the expansion coefficient. One gets
\begin{eqnarray}\label{proj}
[j^{n-1}(\alpha_1',v-1,J_1')jJ|\}j^{n}\beta_i v J]=\beta_i[j^{n-1}(\alpha_1',v-1,J_1')jJ|\}j^{n} v (\alpha_1'J_1') J],
\end{eqnarray}
and
\begin{eqnarray}
\nonumber &&[j^{n-1}(\alpha_1,v-1,J_1)jJ|\}j^{n} v (\alpha_1'J_1') J] [j^{n-1}(\alpha_1',v-1,J_1')jJ|\}j^{n} v (\alpha_1'J_1') J]\\
\nonumber&=&\sum_{\beta_i}\beta_i[j^{n-1}(\alpha_1,v-1,J_1)jJ|\}j^{n}\beta_i v J] [j^{n-1}(\alpha_1',v-1,J_1')jJ|\}j^{n} v (\alpha_1'J_1') J]\\
&=&\sum_{\beta_i} [j^{n-1}(\alpha_1,v-1,J_1)jJ|\}j^{n}\beta_i v J] [j^{n-1}(\alpha_1',v-1,J_1')jJ|\}j^{n}\beta_i v J].
\end{eqnarray}

These recursion relations can be used to calculate the values of the one-particle cfp's. As an example, for the configuration $|j^3;I_3\rangle$, 
one has,
\begin{eqnarray}\label{cfp3}
\nonumber&&\sum_{\nu_3}[j^2 (\nu=2, I)j I_3 |\}j^{3}, \nu_3, I_3] [j^2 (\nu=2, J)j I_3 |\}j^{3}, \nu_3, I_3]
\\\nonumber&=&\frac{1}{3}\delta_{IJ}+\frac{2}{3}(-1)^{I+J}\sqrt{(2I+1)(2J+1)}\left\{\begin{array}{ccc}
j&j&J\\
I_3&j&I\\
\end{array}\right\}[j j I |\}j^{2}, \nu=2, I] [j
j J |\}j^{2}, \nu=2, J ]\\
&=&\frac{1}{3}\delta_{IJ}+\frac{2}{3}(-1)^{I+J}\sqrt{(2I+1)(2\lambda+1)}\left\{\begin{array}{ccc}
j&j&J\\
I_3&j&I\\
\end{array}\right\},
\end{eqnarray}
where we have the relation $\left[j j I |\}j^{2}, \nu=2, I \right]=1$ for even $I$ values. In deriving this relation we have also applied Eq. (19.31) of Ref. \cite{Talmi93}. Above relation can also be derived by using the modified Redmond recursion relation of Ref. \cite{Zam06}.
For shells with $j\leq9/2$, this equation determines the explicit expressions for the cfp's since all the $v=3$ states are unique.

The relations mentioned above are valid in general for all shells. In the following sections we will apply them to analyze the expressions for the Hamiltonian matrix elements. Before that, we will show that they can be used to derive a simple proof of Eq. (\ref{cfp1}).
As in Ref. \cite{Qi11}, we start from  the unique state $|j^5, v=5, J=j\rangle$. It can be easily constructed as the tensor product of any $n=4$ state times the single particle. 
For the principal parent one can take $|\alpha_1'J_1'\rangle=|j^{4},v=4, J_1'=0\rangle$. Since $J_2$ can only take the value $J_2=0$,  in this case we have
\begin{eqnarray}
\nonumber&&5[j^{4}(v=4,\alpha_1,J_1)jJ|\}j^{5}, v=5,
J=j] [j^{4}(v_1'=4,J_1'=0)jJ|\}j^{5}, v=5,
J=j]\\
&=&v\nonumber R'(j,v-1,\alpha_1J_1;v(\alpha_1'J_1') J)  R'(j,v-1,\alpha_1'J_1';v(\alpha_1'J_1') J)\\
&=&\left[\delta_{\alpha_1'\alpha_1}\delta_{J_1'J_1}- (v-1)\frac{\sqrt{2J_1+1} (2j-3)}{(2j+1)(2j-5)}\right] R(j,v=3,j;v=4,\alpha_1 J_1)\nonumber\\
&=&\left[\delta_{\alpha_1'\alpha_1}\delta_{J_1'J_1}-\frac{3\sqrt{2J_1+1}}{5}\right][j^3 (\nu_3=3, I_3=j)j I |\}j^{4},\alpha_1,\nu=4, J_1],
\end{eqnarray}
which is valid for all the $v=4$ states $|\alpha_1J_1\rangle$. For $J_1=I$, it is equivalent to the special relation of Eq. (\ref{cfp1}). For the $j=9/2$ shell we have $[j^{4}(v_1'=4,J_1'=0)jJ|\}j^{5}, v=5,
J=j]=\sqrt{2}/5$.

\section{The non-diagonal matrix elements}

A four-particle state $|j^4;I\rangle$ can be written
in terms of three-particle states with the help of 
one-particle cfp's as,
\begin{eqnarray}
|j^4,\alpha,v,I\rangle= \sum_{v_3,\alpha_3,I_3} [j^3(\alpha_3, v_3, I_3)jI|\}j^4, \alpha, v,I] |j^3(\alpha_3, v_3, I_3)j;I\rangle,
\end{eqnarray}
where the state on the right-hand side results from the coupling of
a three-particle state $\alpha_3$ with the
last particle (denoted as $j$).
For the intermediate seniority of the three-particle states one has $|v-v_3|=1$. 
Thus for $v=4$ the seniority of the three-particle states can only take the value $v_3=3$. In particular, for the $j=9/2$ shell of concern, the index $\alpha_3$ can be omitted since all the three-particle states can be uniquely specified by the quantum numbers $I$ and $v$.
The Hamiltonian matrix element can be calculated  in terms of one-particle cfp's as,
\begin{eqnarray}
M^I_{vv'}(J)=\sum_{v_3v_3'I_3}
[j^3(v_3I_3)jI|\}j^4,v,I][j^3(v'_3=3I_3)jI|\}j^4,v',I]M^{I_3}_{v_3v_3'}(J),
\end{eqnarray}
where the three-particle matrix element can be
expressed as
\begin{equation}
M^{I_3}_{v_3v_3'}(J)=
[j^2(J)jI_3|\}j^3v_3I_3][j^2(J)jI_3|\}j^3v_3'I_3].
\end{equation}

\subsection{The matrices $M^I_{2\alpha}$}

The non-diagonal matrix element between the $v=2$ and $v=4$ states is
%\begin{widetext}
\begin{subequations}
\begin{eqnarray}\label{M2a}
\nonumber M^I_{2\alpha}(J)&=&\sum_{\nu_3=1,3 I_3} [j^3 (\nu_3, I_3 )j I
|\}j^{4} \nu=2 I ] [j^3 (\nu_3'=3, I_3 )j I |\}j^{4},\alpha,
\nu=4, I]\\
\nonumber&&\times  [j^2(J)jI_3|\}j^3v_3I_3][j^2(J)jI_3|\}j^3,v_3'=3,I_3]\\
\nonumber&=&\sum_{I_3}[j^3 (\nu_3=3, I_3 )j I
|\}j^{4} \nu=2 I] [j^3 (\nu_3'=3, I_3 )j I |\}j^{4},\alpha,
\nu=4, I]\\
 &&\times [j^2(J)jI_3|\}j^3,v_3=3,I_3][j^2(J)jI_3|\}j^3,v_3'=3,I_3]\\
&& +[j^3 (\nu_3=1, I_3=j)j I
|\}j^{4} ,\nu=2, I][j^3 (\nu_3'=3, I_3=j)j I |\}j^{4},\alpha,\nu=4, I]M^{I_3=j}_{13}(J),~~~~~~~~~
\end{eqnarray}
\end{subequations}
%\end{widetext}
where $M^{I_3=j}_{13}(J)=[j^2(J)jI_3|\}j^3,v_3=1,I_3=j] [j^2(J)jI_3|\}j^3,v_3'=3,I_3=j]$.

From Eq. (19.31) of Ref. \cite{Talmi93}, one readily gets,
\begin{eqnarray}\label{M2a1}
\nonumber[j^3 (\nu_3, I_3 )j I
|\}j^{4} \nu=2 I][j^2 (J )j I_3 |\}j^{3} \nu_3 I_3
]\\
=\sqrt{\frac{2J+1}{2I+1}}[j^2
(\nu=2, I)j I_3 |\}j^{3} \nu_3 I_3][j^3
(\nu_3 I_3 )j J |\}j^{4},\nu=2,J].~~~~
\end{eqnarray}

By applying this relation and Eq. (\ref{cfp3}), the first part of $M^I_{2\alpha}$ becomes,
\begin{widetext}
\begin{eqnarray}
\nonumber M^I_{2\alpha}(J;a)&=&\sqrt{\frac{2J+1}{2I+1}}\sum_{\nu_3
I_3}\left[\frac{1}{3}\delta_{IJ}+\frac{2}{3}\sqrt{(2J+1)(2I+1)}\left\{\begin{array}{ccc}
j&j&J\\
I_3&j&I\\
\end{array}\right\}\right]\\
\nonumber && \times [j^3(\nu_3 I_3 )j J |\}j^{4},\nu=2, J][j^3
(\nu_3, I_3 )j I |\}j^{4} \nu=4,\alpha, I]\\
\nonumber &&- [j^2 (\nu=2, I)j I_3 |\}j^{3}, \nu_3=1, I_3=j] [j^2 (\nu=2, J)j I_3 |\}j^{3}, \nu_3=1, I_3=j]\\
\nonumber&&\times[j^3(\nu_3 I_3=j )j J |\}j^{4},\nu=2, J][j^3
(\nu_3, I_3=j )j I |\}j^{4} \nu=4,\alpha, I]\\
\nonumber&=&\frac{5}{6}
\sqrt{\frac{2J+1}{2I+1}}
\left[j^4 (\nu=2, J )j I_5 |\}j^{5},\nu_5=3,I_5=j
\right]\left[j^4 (\nu=4,\alpha, I )j I_5 |\}j^{5}, \nu_5=3, I_5=j \right]\\
\nonumber&&- \sqrt{\frac{2J+1}{2I+1}}[j^2 (\nu=2, I)j I_3 |\}j^{3}, \nu_3=1, I_3=j] [j^2 (\nu=2, J)j I_3 |\}j^{3}, \nu_3=1, I_3=j]\\
&&\times[j^3(\nu_3 I_3=j )j J |\}j^{4},\nu=2, J][j^3
(\nu_3, I_3=j )j I |\}j^{4} \nu=4,\alpha, I].
\end{eqnarray}
Combining above equation with the second part of $M^I_{2\alpha}(J)
$, we get
\begin{eqnarray}
\nonumber M^I_{2\alpha}(J)&=&\frac{5}{6}
\sqrt{\frac{2J+1}{2I+1}}
\left[j^4 (\nu=2, J )j I_5 |\}j^{5},\nu_5=3,I_5=j
\right][j^4 (\nu=4,\alpha, I )j I_5 |\}j^{5}, \nu_5=3, I_5=j]\\
&& +\frac{4}{3}[j^3 (\nu_3=3, I_3=j)j I |\}j^{4},\alpha,\nu=4, I]   [j^3 (\nu_3=1, I_3=j)j I
|\}j^{4} ,\nu=2, I]M^{I_3=j}_{13}(J).~~~~~~~
\end{eqnarray}
\end{widetext}

This expression can be further simplified by noting that the following relation should hold, 
\begin{eqnarray}
 [j^3 (\nu_3=1, I_3=j)j I
|\}j^{4} ,\nu=2, I]=\sqrt{\frac{v(2j+3-n-v)}{n(2j+3-2v)}}=\sqrt{\frac{3}{8}},
\end{eqnarray}
and
\begin{eqnarray}
[j^4 (\nu=4,\alpha, I )j I_5 |\}j^{5}, \nu_5=3, I_5=j]=\frac{\sqrt{2I+1}}{5}[j^3 (\nu_3=3, I_3=j)j I |\}j^{4},\alpha,\nu=4, I].~~~
\end{eqnarray}
These are derived from Eq. (\ref{sun}) and Eq. (19.31) of Ref. \cite{Talmi93}, respectively. Finally we have
\begin{eqnarray}
\nonumber M^I_{2\alpha}(J)&=&\left[\frac{\sqrt{2J+1}}{6}
[j^4 (\nu=2, J )j I_5 |\}j^{5},\nu_5=3,I_5=j
]
 +\sqrt{\frac{2}{3}}M^{I_3=j}_{13}(J)\right]\\
\nonumber&&\times[j^3 (\nu_3=3, I_3=j)j I |\}j^{4},\alpha,\nu=4, I]\\
&=&\left[\frac{2}{3}\sqrt{\frac{2J+1}{10}}[j^2(J)jI_3|\}j^3,v_3=3,I_3=j]\right][j^3 (\nu_3=3, I_3=j)j I |\}j^{4},\alpha,\nu=4, I].~~~~~~~
\end{eqnarray}
The term in the bracket on the right-hand side of this equation defines the seniority conservation condition of the interaction \cite{Talmi93,Isacker08,Qi10a,Rowe03}.

\subsection{The matrices $M^I_{\alpha_1\alpha_2}$}

For the matrix elements involving the two $v=4$ states it is,
%\begin{widetext}
\begin{eqnarray}
\nonumber  M^I_{\alpha_1\alpha_2}(J)&=&\sum_{v_3,v_3',I_3} [j^3 (\nu_3, I_3 )j I
|\}j^{4}, \alpha_1, \nu=4, I ] [j^3 (\nu_3', I_3 )j I |\}j^{4},\alpha_2,
\nu=4, I ]M^{I_3}_{v_3v_3'}(J) \\
\nonumber  &=&\sum_{I_3} [j^3 (\nu_3=3, I_3 )j I
|\}j^{4}, \alpha_1, \nu=4, I ] [j^3 (\nu_3=3, I_3 )j I |\}j^{4},\alpha_2,
\nu=4, I ] \\
\nonumber
&&\times [j^2(J)jI_3|\}j^3,v_3=3,I_3]^2\\
\nonumber &=&\sum_{v_3'=1,3I_3} [j^3 (\nu_3=3, I_3 )j I
|\}j^{4}, \alpha_1, \nu=4, I ] [j^3 (\nu_3=3, I_3 )j I |\}j^{4},\alpha_2,
\nu=4, I ] \\
\nonumber&&\times [j^2(J)jI_3|\}j^3,v_3',I_3]^2\\
\nonumber&&-[j^3 (\nu_3=3, I_3=j )j I
|\}j^{4}, \alpha_1, \nu=4, I ] [j^3 (\nu_3=3, I_3=j )j I |\}j^{4},\alpha_2,
\nu=4, I ]\\
&&\times  [j^2(J)jI_3|\}j^3,v_3=1,I_3=j]^2.~~~~~
\end{eqnarray}
By applying the special relation of Eq. (\ref{cfp3}), one gets
\begin{subequations}
\begin{eqnarray}
\nonumber M^I_{\alpha_1\alpha_2}(J)&=&\sum_{v_3I_3} [j^3 (\nu_3, I_3 )j I
|\}j^{4}, \alpha_1, \nu=4, I ] [j^3 (\nu_3, I_3 )j I |\}j^{4},\alpha_2,
\nu=4, I ]\\
&&\times\left[ \frac{1}{3}+\frac{2}{3}(2J+1)\left\{\begin{array}{ccc}
I_3&j&J\\
j&j&J\\
\end{array}\right\}\right]\\
\nonumber&& -[j^3 (\nu_3=3, I_3=j )j I
|\}j^{4}, \alpha_1, \nu=4, I ] [j^3 (\nu_3=3, I_3=j )j I |\}j^{4},\alpha_2,
\nu=4, I]\\
&&\times[j^2(J)jI_3|\}j^3,v_3=1,I_3=j]^2.~~~~~~
\end{eqnarray}
\end{subequations}
%\end{widetext}
For the special case $I=J$, the first part of this equation can be simplified as \cite{Zam08}
\begin{eqnarray}
\nonumber M^I_{\alpha_1\alpha_2}(J=I;a)&=&\frac{5}{6}\sum_{v_5=3,5}
\left[j^4 (\nu=4,\alpha_1, I)j I_5 |\}j^{5},\nu_5,I_5=j
\right]\left[j^4 (\nu=4,\alpha_2, I )j I_5 |\}j^{5}, \nu_5, I_5=j \right]\\
\nonumber&&+\frac{1}{6}\delta_{\alpha_1\alpha_2}\\
\nonumber&=&\frac{11(2I+1)}{60}[j^3 (\nu_3=3, I_3=j )j I
|\}j^{4}, \alpha_1, \nu=4, I ]\\
&&\times[j^3 (\nu_3=3, I_3=j )j I
|\}j^{4}, \alpha_2, \nu=4, I ]+\frac{1}{6}\delta_{\alpha_1\alpha_2}.~~~~~~~
\end{eqnarray}
Thus, the matrix elements $M^I_{\alpha_1\alpha_2}(J=I)$ can be simplified as
\begin{eqnarray}
\nonumber M^I_{\alpha_1\alpha_2}(J=I)&=&\frac{(2I+1)}{6}[j^3 (\nu_3=3, I_3=j )j I
|\}j^{4}, \alpha_1, \nu=4, I ]\\
&&\times[j^3 (\nu_3=3, I_3=j )j I
|\}j^{4}, \alpha_2, \nu=4, I ]+\frac{1}{6}\delta_{\alpha_1\alpha_2}.~~
\end{eqnarray}
Immediately one realizes that for $I=J$, the matrix elements of $M^I(J)$ are proportional to $[j^3 (\nu_3=3, I_3=j )j I
|\}j^{4}, \alpha, \nu=4, I ]$ (and $\left[j^4 (\nu=4,\alpha, I)j I_5 |\}j^{5},\nu_5,I_5=j
\right]$ with $v_5=3$ and 5).

We did not get a simple expression for the matrix elements of $M^I(J)$ with $J\neq I$ by directly applying the recursion relations mentioned in Section II. But we found that one can do so by exploring the special properties of the $v=4$ states.
One may write the special $v=4$ state of concern as a combination of an arbitrary set of $v=4$ states as \cite{Zam08}
\begin{equation}
|j^4,a,v=4,I\rangle=\alpha|j^4,\alpha_1,v=4,I\rangle+\beta|j^4,\alpha_2,v=4,I\rangle,
\end{equation}
where the amplitudes are denoted by  $\alpha$ and $\beta$. One should keep in mind that it is trivial to construct a special $v=4$ state that satisfies $H^I_{2a}=M^I_{2a}(J)=0$ by taking into account the fact that in $j=9/2$ shell there is only one seniority conservation condition.
For such a state, immediately we can have,
\begin{eqnarray}
\nonumber \frac{H^I_{2\alpha_1}}{H^I_{{2\alpha_{2}}}} &=&\frac{M^I_{2\alpha_1}(J)}{M^I_{{2\alpha_{2}}}(J)} = -\frac{\beta}{\alpha}=\frac{[j^3 (\nu_3=3, I_3=j)j I |\}j^{4},\alpha_1,\nu=4, I] }{[j^3 (\nu_3=3, I_3=j)j I |\}j^{4},\alpha_2,\nu=4, I] },
\end{eqnarray}
which do not depend on the values of $J$. If the state $|j^4,a,v=4,I\rangle$ thus constructed is an eigenstate of any Hamiltonian $H$, we should also have
\begin{equation}\label{Mab}
\frac{H^I_{\alpha_1\alpha_1}-H^I_{\alpha_{2}\alpha_{2}}}{H^I_{\alpha_1\alpha_2}}
= \frac{M^I_{\alpha_1\alpha_1}(J)-M^I_{\alpha_{2}\alpha_{2}}(J)}{M^I_{\alpha_1\alpha_2}(J)}
= \left[  \frac{\alpha}{\beta}-\frac{\beta}{\alpha}  \right].
\end{equation}
This is indeed the case for the special case of $J=I$ where 
we have $M^I_{ab}(I)=0$ and \cite{Zam08}
\begin{eqnarray}
\nonumber&&\left[j^4 (\nu=4,a, I )j I_5 |\}j^{5}, \nu_5=5, I_5=j \right]\\
\nonumber&=&\left[j^4 (\nu=4,a, I )j I_5 |\}j^{5}, \nu_5=3, I_5=j \right]\\
\nonumber&=&[j^3 (\nu_3=3, I_3=j)j I |\}j^{4},a,\nu=4, I]\\
&=&0.
\end{eqnarray}
These relations imply that the special $v=4$ state is an eigenstate of the seniority-nonconserving interaction of $V_I$. The $v=4$ state orthogonal to it can be explicitly constructed as
\begin{eqnarray}
[j^3 (\nu_3=3, I_3)j I |\}j^{4},b,\nu=4, I]=[j^3 (\nu_3=3, I_3)j I |\}j^{4},\nu=4(v_3'=3,I_3'=j) I].
\end{eqnarray}
The one-body cfp's of the special $I=4$ and $6$ states can thus be calculated through the help of symbolic calculations. These are listed in Table I and II, respectively. It is thus found that  the non-diagonal matrix elements involving the special $v=4$ states indeed vanish.
For instance, one has
\begin{eqnarray}
\nonumber M^{I=4}_{ab}(J=2)&=&\left[-\frac{76\sqrt{221}}{3\sqrt{105}}\frac{5}{18}+\frac{95\sqrt{221}}{2\sqrt{105}}\frac{52}{99} +\frac{176\sqrt{1785}}{9\sqrt{13}}\frac{17}{99}-\frac{184\sqrt{1785}}{10\sqrt{13}}\frac{10}{33}\right]\frac{1}{1727}\\
&=&0,
\end{eqnarray}
and
\begin{eqnarray}
\nonumber M^{I=4}_{ab}(J=8)&=&\left[\frac{95\sqrt{221}}{2\sqrt{105}}\frac{238}{715}
+\frac{176\sqrt{1785}}{9\sqrt{13}}\frac{228}{715}-\frac{184\sqrt{1785}}{10\sqrt{13}}\frac{323}{858}
+\frac{38\sqrt{32851}}{3\sqrt{255}}\frac{3}{13}\right.\\
\nonumber &&+\left.\frac{30\sqrt{455}}{\sqrt{51}}\frac{209}{390}\right]\frac{1}{1727}\\
&=&0.
\end{eqnarray}
The calculated diagonal matrix elements are given in Table III.

\begin{table}
  \centering
  \caption{One-particle cfp's $[j^3(v_3I_3)jI|\}j^4,\alpha,I]$ for states $|j^4,a,v=4,I=4\rangle$ and $|j^4,b,v=4,I=4\rangle$.}
\begin{ruledtabular}
  \begin{tabular}{ccc}
$I_3$&$a$&$b$\\
\hline
3/2&    $\displaystyle-\sqrt{\frac{1547}{1727*60}} $   &   $\displaystyle\frac{59}{3\sqrt{1727}}$   \\
5/2&      $\displaystyle\frac{8\sqrt{17}}{\sqrt{210*1727}}$  &  $\displaystyle-\frac{19\sqrt{13}}{3\sqrt{2*1727}}$         \\
7/2&    $\displaystyle-\frac{19\sqrt{51}}{2\sqrt{7*1727}}$    &   $\displaystyle-\frac{\sqrt{65}}{3\sqrt{1727}}$    \\
9/2&    0    &$\displaystyle\frac{\sqrt{1727}}{33\sqrt{13}}$   \\
11/2&  $\displaystyle\frac{11\sqrt{21}}{3\sqrt{1727}}$     &    $\displaystyle\frac{16\sqrt{85}}{3\sqrt{13*1727}}$          \\
13/2&  $\displaystyle-\frac{8\sqrt{255}}{5\sqrt{1727}}$       &    $\displaystyle\frac{23\sqrt{7}}{2\sqrt{13*1727}}$     \\
15/2&  $\displaystyle\frac{13\sqrt{133}}{\sqrt{510*1727}}$     &   $\displaystyle\frac{76\sqrt{19}}{3\sqrt{26*1727}}$         \\
17/2&   $\displaystyle\frac{12\sqrt{7}}{\sqrt{17*1727}}$        &  $\displaystyle-\frac{5\sqrt{65}}{2\sqrt{3*1727}}$      \\
  \end{tabular}
  \end{ruledtabular}
\end{table}

\begin{table}
  \centering
  \caption{One-particle cfp's $[j^3(v_3I_3)jI|\}j^4,\alpha,I]$ for states $|j^4,a,v=4,I=6\rangle$ and $|j^4,b,v=4,I=6\rangle$.}
\begin{ruledtabular}
  \begin{tabular}{ccc}
$I_3$&$a$&$b$\\
\hline
3/2&    $\displaystyle \sqrt{\frac{2261}{715*281}} $   &   $\displaystyle -\frac{46\sqrt{3}}{3\sqrt{143*281}}$  \\
5/2&      $\displaystyle -\frac{4\sqrt{646}}{\sqrt{385*281}}$  &  $\displaystyle-\frac{43}{\sqrt{66*281}}$      \\
7/2&    $\displaystyle -\frac{45\sqrt{323}}{\sqrt{6006*281}}$    &   $\displaystyle\frac{2\sqrt{10}}{3\sqrt{143*281}}$   \\
9/2&    0    & $\displaystyle\frac{\sqrt{281}}{3\sqrt{286}}$     \\
11/2&  $\displaystyle -\frac{55\sqrt{19}}{\sqrt{4862*281}}$     &    $\displaystyle\frac{13\sqrt{210}}{3\sqrt{143*281}}$            \\
13/2&  $\displaystyle-\frac{42\sqrt{19}}{5\sqrt{715*281}}$       &    $\displaystyle\frac{13\sqrt{119}}{4\sqrt{429*281}}$        \\
15/2&  $\displaystyle\frac{507\sqrt{3}}{\sqrt{230945*281}}$     &   $\displaystyle-\frac{43\sqrt{7}}{\sqrt{143*281}}$          \\
17/2&   $\displaystyle\frac{70\sqrt{21}}{\sqrt{2431*281}}$        &  $\displaystyle\frac{177\sqrt{19}}{4\sqrt{715*281}}$      \\
21/2&    $\displaystyle\frac{1320}{\sqrt{46189*281}}$   &  $\displaystyle-\frac{77\sqrt{7}}{\sqrt{8580*281}}$    \\
  \end{tabular}
  \end{ruledtabular}
\end{table}

It is thus seen that the diagonal matrix elements can be rewritten as
\begin{eqnarray}
M^{I}_{bb}(J)&=&M^{I}_{aa}(J)+C_J[j^3 (\nu_3=3, I_3=j)j I |\}j^{4},b,\nu=4, I]^2,
\end{eqnarray}
where $C_J=$ $-13/6$, $3/2$, $13/6$ and $-3/2$ for $J=2$, 4, 6 and 8, respectively. For an arbitrary set of $v=4$ states
\begin{eqnarray}
|j^4,\alpha_1,v=4,I\rangle&=&\alpha|j^4,a,v=4,I\rangle-\beta|j^4,b,v=4,I\rangle,
\end{eqnarray}
and
\begin{eqnarray}
|j^4,\alpha_2,v=4,I\rangle&=&\beta|j^4,a,v=4,I\rangle+\alpha|j^4,b,v=4,I\rangle,
\end{eqnarray}
by applying Eq. (\ref{proj}), we have
\begin{eqnarray}
[j^3 (\nu_3=3, I_3=j)j I |\}j^{4},\alpha_1,\nu=4, I]=-\beta[j^3 (\nu_3=3, I_3=j)j I |\}j^{4},b,\nu=4, I],
\end{eqnarray}
and
\begin{eqnarray}
[j^3 (\nu_3=3, I_3=j)j I |\}j^{4},\alpha_2,\nu=4, I]=\alpha[j^3 (\nu_3=3, I_3=j)j I |\}j^{4},b,\nu=4, I].
\end{eqnarray}
Thus, the matrix elements corresponding to an arbitrary set of $v=4$ states can be rewritten as
\begin{eqnarray}
\nonumber M^I_{\alpha_1\alpha_1}(J)&=&\alpha^2M^I_{aa}(J) + \beta^2M^I_{bb}(J)\\
\nonumber&=&M^I_{aa}(J)+C_J\left(\frac{[j^3 (\nu_3=3, I_3=j)j I |\}j^{4},\alpha_1,\nu=4, I]}{[j^3 (\nu_3=3, I_3=j)j I |\}j^{4},b,\nu=4, I]}\right)^2\\
\nonumber&&\times[j^3 (\nu_3=3, I_3=j)j I |\}j^{4},b,\nu=4, I]^2\\
&=&M^I_{aa}(J)+C_J[j^3 (\nu_3=3, I_3=j)j I |\}j^{4},\alpha_1,\nu=4, I]^2,
\end{eqnarray}
and
\begin{eqnarray}
M^I_{\alpha_1\alpha_2}(J)&=&\alpha\beta\left[M^I_{aa}(J)-M^I_{bb}(J)\right]\\
\nonumber&=&C_J[j^3 (\nu_3=3, I_3=j)j I |\}j^{4},\alpha_1,\nu=4, I]}{[j^3 (\nu_3=3, I_3=j)j I |\}j^{4},\alpha_2,\nu=4, I].
\end{eqnarray}

\begin{table}
  \centering
  \caption{Diagonal matrix elements of the matrices $M^I(J)$ for the special $v=4$ states $|j^4,a,v=4,I\rangle$ and $|j^4,b,v=4,I\rangle$.}
\begin{ruledtabular}
  \begin{tabular}{ccccc}
&$J=2$& $J=4$ & $J=6$ & $J=8$\\
\hline
$M^{I=4}_{aa}$&34/99&1/6&13/90&19/55\\
$M^{I=4}_{bb}$&47/594&50/143&607/1485&697/4290\\
$M^{I=6}_{aa}$&19/66&2/13&1/6&56/143\\
$M^{I=6}_{bb}$&61/1188&545/1716&479/1188&391/1716\\
  \end{tabular}
  \end{ruledtabular}
\end{table}

One can also show that the following relations hold,
\begin{eqnarray}
 \frac{M^I_{\alpha_1\alpha_1}(J)-M^I_{aa}(J)}{M^I_{\alpha_1\alpha_2}(J)}
= \frac{M^I_{bb}(J)-M^I_{\alpha_2\alpha_2}(J)}{M^I_{\alpha_1\alpha_2}(J)}
=\frac{M^I_{2\alpha_1}(J)}{M^I_{{2\alpha_{2}}}(J)} = -\frac{\beta}{\alpha} ,
\end{eqnarray}
and
\begin{eqnarray}
 \frac{M^I_{\alpha_2\alpha_2}(J)-M^I_{aa}(J)}{M^I_{\alpha_1\alpha_2}(J)}= \frac{M^I_{bb}(J)-M^I_{\alpha_1\alpha_1}(J)}{M^I_{\alpha_1\alpha_2}(J)}
=\frac{M^I_{2\alpha_2}(J)}{M^I_{{2\alpha_{1}}}(J)} = -\frac{\alpha}{\beta}.
\end{eqnarray}
These are equivalent to the special relation of Eq. (\ref{Mab}).

\subsection{Two-particle cfp's}

The advantage of applying one-particle cfp's is that, for the problem of concern, the Hamiltonian matrix elements can be re-expressed in simple ways. 
But for a system in general, the Hamiltonian matrix elements may be calculated in a more straightforward way with the help of two-particle cfp's.
The matrix elements of  $M^I(J)$ can be expressed in terms of two-particle cfp's as \cite{Talmi93,Talmi10,Isacker08,Qi10},
\begin{eqnarray}
\nonumber M^I_{\alpha\beta}(J) &=&\sum_{\alpha_{n-2}I_{n-2}}[j^{n-2}(\alpha_{n-2}I_{n-2})j^2(J)I|\}j^n\alpha I] [j^{n-2}(a_{n-2}I_{n-2})j^2(J)I|\}j^n\beta I],
\end{eqnarray}
where the summation runs over all possible $j^{n-2}$ states.

The two-particle cfp's can be constructed within the principal parent scheme \cite{Talmi93, Isacker08} and be expressed in closed forms in terms of $9j$ symbols. 
For four identical nucleons in a single-$j$ shell, the state can be written as the tensor product of two-particle states as $|j^2(J_{\alpha})j^2(J_{\beta});I\rangle$. The overlap between such states (which takes into account the Pauli principle) is~\cite{Liotta81,Qi10,Zhao03b},
\begin{eqnarray}
\nonumber A^j_I(J_{\alpha}J_{\beta};J_{\alpha}'J_{\beta}')&=&\langle j^2(J_{\alpha})j^2(J_{\beta});I|j^2(J'_{\alpha})j^2(J'_{\beta});I\rangle\\
\nonumber &=&\delta_{J_{\alpha}J_{\alpha}'}\delta_{J_{\beta}J_{\beta}'}+(-1)^{I}\delta_{J_{\alpha}J_{\beta}'}\delta_{J_{\beta}J_{\alpha}'}-4\hat{J}_{\alpha}\hat{J}_{\beta}\hat{J}_{\alpha}'\hat{J}_{\beta}'
\left\{
\begin{array}{ccc}
j&j&J_{\alpha}\\
j&j&J_{\beta}\\
J_{\alpha}'&J_{\beta}'&I
\end{array}
\right\},
\end{eqnarray}
where $\hat{J}=\sqrt{2J+1}$.
For a given angular momentum $I$, the seniority $v=2$ state is unique and can be written as (see, e.g., Ref. \cite{Isacker08})
\begin{equation}
|j^4,v=2,I\rangle = \mathcal{N}_{0I}|j^2(0)j^2(J=I);I\rangle,
\end{equation}
where $\mathcal{N}_{0I}=[A^j_I(0I;0I)]^{-1/2}$ is the normalization factor and $I$ can take even values with $0<I\leq2j-1$.
For the corresponding two-particle cfp's we have,
\begin{equation}
[j^2(J)j^2(J')I|\}j^4,v=2,I] = \sqrt{\frac{2}{n(n-1)}}\mathcal{N}_{0I}A^j_I(JJ';0I).
\end{equation}

One of the seniority $v=4$ states can be written as
\begin{eqnarray}
|j^4[J_{\alpha}J_{\beta}],v=4,I\rangle = \mathcal{N}_{J_{\alpha}J_{\beta}}|j^2(J_{\alpha})j^2(J_{\beta});I\rangle- \mathcal{N}_{J_{\alpha}J_{\beta}}\langle|j^2(J_{\alpha})j^2(J_{\beta})I|j^4,v=2,I\rangle|j^4,v=2,I\rangle,~~~~~~~
\end{eqnarray}
where $J_{\alpha}$ and $J_{\beta}$ are the so-called principal parents. 
The two-particle cfp of above $v=4$ state are given as,
\begin{eqnarray}
\nonumber [j^2(K)j^2(K')I|\}j^4[J_{\alpha}J_{\beta}],v=4,I] \\
= \sqrt{\frac{2}{n(n-1)}}\mathcal{N}_{J_{\alpha}J_{\beta}}\left[A^j_I(KK';J_{\alpha}J_{\beta})-\mathcal{N}^2_{0I}A^j_I(J_{\alpha}J_{\beta};0I)A^j_I(KK';0I)\right].
\end{eqnarray}
The other $v=4$ state can be constructed through the schmidt orthogonalization procedure in a similar way.

For $I=4$, one can construct the $v=2$ state as $|j^4[J_{\alpha}=0J_{\beta}=4],v=2,I\rangle$ and the first $v=4$ state as $|j^4[J_{\alpha}=2J'_{\beta}=2],v=4,I\rangle$.
The upper part of the symmetric matrix $M$ is thus calculated to be
\begin{eqnarray}
M^{I=4}(J=2)=\frac{1}{6}\left(
\begin{array}{ccc}
\displaystyle\frac{67}{99}&-\displaystyle\frac{\sqrt{182}}{99\sqrt{163}}&-\displaystyle\frac{10\sqrt{510}}{99\sqrt{163}}\\[1em]
&\displaystyle\frac{33161}{16137}&\displaystyle-\frac{10\sqrt{7735}}{5379\sqrt{3}}\\[1em]
&&\displaystyle\frac{2584}{5379}\\
\end{array}\right),
\end{eqnarray}
\begin{equation}
M^{I=4}(J=4)=\frac{1}{6}\left(\begin{array}{ccc}
                \displaystyle \frac{746}{715} &\displaystyle \frac{7\sqrt{14}}{11\sqrt{2119}}  &\displaystyle \frac{70\sqrt{510}}{143\sqrt{163}} \\[1em]
                     &\displaystyle \frac{1800}{1793}  &\displaystyle \frac{10\sqrt{1785}}{1793\sqrt{13}} \\[1em]
                     &    &\displaystyle \frac{48809}{23309}  \\
              \end{array}\right),
\end{equation}
\begin{equation}
M^{I=4}(J=6)=\frac{1}{6}\left(\begin{array}{ccc}
                \displaystyle \frac{1186}{495} &\displaystyle -\frac{31\sqrt{182}}{495\sqrt{163}}  &\displaystyle -\frac{62\sqrt{170}}{33\sqrt{489}} \\[1em]
                     &\displaystyle \frac{70382}{80685}  &\displaystyle \frac{10\sqrt{7735}}{5379\sqrt{3}} \\[1em]
                     &    &\displaystyle \frac{65809}{26895}  \\
              \end{array}\right),
\end{equation}
and
\begin{equation}
M^{I=4}(J=8)=\frac{1}{6}\left(\begin{array}{ccc}
                \displaystyle \frac{918}{715} &\displaystyle \frac{17\sqrt{14}}{55\sqrt{2119}}  &\displaystyle \frac{34\sqrt{510}}{143\sqrt{163}} \\[1em]
                     &\displaystyle \frac{18547}{8965}  &\displaystyle -\frac{10\sqrt{1785}}{1793\sqrt{13}} \\[1em]
                     &    &\displaystyle \frac{114066}{116545}  \\
              \end{array}\right),
\end{equation}
from which one can get
\begin{equation}
\frac{M^I_{2\alpha_1}(J)}{M^I_{{2\alpha_{2}}}(J)} = -\frac{\beta}{\alpha} = \frac{\sqrt{182}} {10\sqrt{510}},
\end{equation}
\begin{eqnarray}
[j^2(J_{\alpha=2})j^2(J_{\beta}=2)I|\}j^4,a,v=4,I] =\alpha[j^2(J_{\alpha}=2)j^2(J_{\beta}=2)I|\}j^4[J_{\alpha}J_{\beta}],v=4,I],
\end{eqnarray}
and
\begin{equation}
\frac{M^I_{\alpha_1\alpha_1}(J)-M^I_{\alpha_{2}\alpha_{2}}(J)}{M^I_{\alpha_1\alpha_2}(J)}
= \left[  \frac{\alpha}{\beta}-\frac{\beta}{\alpha}  \right]=-\frac{25409}{10\sqrt{23205}}.
\end{equation}
These two relations are valid for all $J$ values of concern. They are sufficient in ensuring that the state $|j^4,a,v=4,I\rangle$ is an eigenstate of any Hamiltonian $H$.

Similarly for $I=6$, one can construct the $v=2$ state as $|j^4[J_{\alpha}=0J_{\beta}=6],v=2,I\rangle$ and the first $v=4$ state as $|j^4[J_{\alpha}=2J'_{\beta}=4],v=4,I\rangle$.
The upper part of the symmetric matrix $M$ is thus calculated to be
\begin{equation}
 M^{I=6}(J=2)=\frac{1}{6}\left(\begin{array}{ccc}
                \displaystyle \frac{34}{99} &\displaystyle \frac{5\sqrt{5}}{99\sqrt{97}}  &\displaystyle -\frac{2\sqrt{2261}}{33\sqrt{291}} \\[1em]
                     &\displaystyle \frac{33049}{19206}  &\displaystyle \frac{5\sqrt{11305}}{3201\sqrt{3}} \\[1em]
                     &    &\displaystyle \frac{1007}{3201}  \\
              \end{array}\right),
\end{equation}
\begin{equation}
 M^{I=6}(J=4)=\frac{1}{6}\left(\begin{array}{ccc}
                \displaystyle \frac{1186}{715} &\displaystyle -\frac{35\sqrt{5}}{143\sqrt{97}}  &\displaystyle \frac{14\sqrt{6783}}{143\sqrt{97}} \\[1em]
                     &\displaystyle \frac{25733}{27742}  &\displaystyle -\frac{5\sqrt{33915}}{13871} \\[1em]
                     &    &\displaystyle \frac{26370}{13871}  \\
              \end{array}\right),
\end{equation}
\begin{equation}
 M^{I=6}(J=6)=\frac{1}{6}\left(\begin{array}{ccc}
                \displaystyle \frac{658}{495} &\displaystyle \frac{31\sqrt{5}}{99\sqrt{97}}  &\displaystyle -\frac{62\sqrt{2261}}{165\sqrt{291}} \\[1em]
                     &\displaystyle \frac{19331}{19206}  &\displaystyle -\frac{5\sqrt{11305}}{3201\sqrt{3}} \\[1em]
                     &    &\displaystyle \frac{7723}{3201}  \\
              \end{array}\right),
\end{equation}
and
\begin{equation}
 M^{I=6}(J=8)=\frac{1}{6}\left(\begin{array}{ccc}
                \displaystyle \frac{1479}{715} &\displaystyle -\frac{17\sqrt{5}}{143\sqrt{97}}  &\displaystyle \frac{34\sqrt{6783}}{715\sqrt{97}} \\[1em]
                     &\displaystyle \frac{65059}{27742}  &\displaystyle \frac{5\sqrt{33915}}{13871} \\[1em]
                     &    &\displaystyle \frac{19026}{13871}  \\
              \end{array}\right).
\end{equation}
For these $I=6$ states we have
\begin{equation}
\frac{M^I_{2\alpha_1}(J)}{M^I_{{2\alpha_{2}}}(J)} = -\frac{\beta}{\alpha} =- \frac{5\sqrt{15}}{6\sqrt{2261}},
\end{equation}
\begin{eqnarray}
[j^2(J_{\alpha=2})j^2(J_{\beta}=4)I|\}j^4,a,v=4,I] =\alpha[j^2(J_{\alpha}=2)j^2(J_{\beta}=4)I|\}j^4[J_{\alpha}J_{\beta}],v=4,I],
\end{eqnarray}
and
\begin{equation}
\frac{M^I_{\alpha_1\alpha_1}(J)-M^I_{\alpha_{2}\alpha_{2}}(J)}{M^I_{\alpha_1\alpha_2}(J)}
= \left[  \frac{\alpha}{\beta}-\frac{\beta}{\alpha}  \right]=\frac{27007}{10\sqrt{33915}}.
\end{equation}
Although the special $v=4$ states $|a\rangle$ cannot be constructed in a straightforward way in terms of one-particle or two-particle cfp's, it is thus noted that these $I=4$ and 6 states have very large overlap with the states $|j^4[J_{\alpha}=2J_{\beta}=2],v=4,I\rangle$ and $|j^4[J_{\alpha}=2J_{\beta}=2],v=4,I\rangle$, respectively. The corresponding overlaps are calculated to be
$\alpha=10\sqrt{255}/\sqrt{25591}$ and $2\sqrt{6783}/\sqrt{27257}$ for the $I$=4 and 6 states.

\section{Summary}
In this paper, the partial conservation of seniority in $j=9/2$
shells is studied within a fully analytic framework with the help of the one-particle and two-particle
cfp's. By using the variety of recursion relations proposed here and from Refs. \cite{Talmi93,Zam06,Red54}, it is shown that the all the relevant diagonal and non-diagonal matrix elements can be expressed in very simple ways in terms of certain one-particle cfp's (c.f., Eqs. (24), (40) \& (41)). This is related to the fact that all $v=3$ and 5 states in $j=9/2$ shells are uniquely defined. It is also found that an alternative proof of the special relation Eq. (1) found in Ref. \cite{Zam08} can be derived by using the recursion relation proposed in this work.

\section*{Acknowlegement}

C.Q. acknowledges the supports from the
Swedish Research Council (VR) under grant No. 621-2010-4723 and from the Swedish National Infrastructure for Computing (SNIC)
at PDC and NSC.

\end{document}